\documentclass{cernrep}
\usepackage{graphicx,here}

\def\be{\begin{equation}}
\def\ee{\end{equation}}

\begin{document}

\title{Effect of cosmic rays on the resonant gravitational
wave detector NAUTILUS at temperature T=1.5 K}

\author{
P. Astone$^1$, D.Babusci$^2$, M. Bassan$^3$, P. Bonifazi$^4$, P. Carelli$^5$,
E. Coccia$^3$\\ S.D'Antonio$^3$,
V. Fafone$^2$, G.Giordano$^2$, A. Marini$^2$, G. Mazzitelli$^2$,
Y. Minenkov$^3$\\ I. Modena$^3$, G. Modestino$^2$,
A. Moleti$^3$, G. V. Pallottino$^6$,\\ G. Pizzella$^7$,
L.Quintieri$^2$, A.Rocchi$^7$, F. Ronga$^2$, R. Terenzi$^4$,
M. Visco$^4$\\
$~$\\
$~$
}

\vskip 0.1 in

\institute{
{\it ${}^{1)}$ Istituto Nazionale di Fisica Nucleare INFN, Rome1}\\
{\it ${}^{2)}$ Istituto Nazionale di Fisica Nucleare INFN, LNF, Frascati}\\
{\it ${}^{3)}$ University of Rome "Tor Vergata" and INFN, Rome2}\\
{\it ${}^{4)}$ IFSI-CNR and INFN, Roma}\\
{\it ${}^{5)}$ University of L'Aquila and INFN, Rome2}\\
{\it ${}^{6)}$ University of Rome "La Sapienza" and INFN, Rome1}\\
{\it ${}^{7)}$ University of Rome "Tor Vergata" and INFN, LNF, Frascati}
}
\date{\today}
\maketitle
\begin{abstract}
The interaction between cosmic rays and the gravitational wave bar detector
NAUTILUS is experimentally studied with the aluminum bar 
at temperature of T=1.5 K. The results are compared with those obtained in 
the previous runs when the bar was at T=0.14 K.
The results of the run at T = 1.5 K are in agreement with 
the thermo-acoustic model;
no large signals at unexpected rate are noticed,
unlike the data  taken in the run at T = 0.14 K.
The observations suggest a larger efficiency in the mechanism of 
conversion of the particle energy into vibrational mode energy
when the aluminum bar is in the superconductive  status.
\end{abstract}
\skip 0.5 in
PACS : 04.80,04.30,96.40.Jj

The gravitational wave (GW) detector NAUTILUS recently recorded
signals due to the passage of cosmic rays (CRs)
\cite{prl,physlet,amaldico}. Several authors \cite{beron}-\cite{bari} 
estimated the
possible acoustic effects due to the passage of particles in a metallic bar.
The mechanism adopted assumes that the
mechanical vibrations originate from local thermal expansion caused by 
warming up due to the energy lost by the particles crossing the material. It
was predicted that for the vibrational energy in the longitudinal fundamental
mode of a metallic bar the following formula would hold:
\be
E=\frac{4}{9\pi}\frac{\gamma^2}{\rho L v^2}
\left(\frac{{\rm d}W}{{\rm d}x}\right)^2
\left(sin\left(\frac{\pi z_0}{L}\right) 
\frac{sin(\pi l_0 cos(\theta_0)/2L)}{\pi Rcos
(\theta_0)/L}\right)^2
\label{desudx}
\ee
where $L$ is the bar length, $R$ the bar radius, $l_0$ the length of the
particle's track inside the bar, $z_0$ the distance of the track midpoint
from one end of the bar, $\theta_0$ the angle between the particle track
 and the axis of the bar, 
$E$ the energy of the excited vibration mode, dW/dx
the energy loss
of the particle in the bar, $\rho$ the density, $v$ the sound velocity in the
material and $\gamma$ is the Gr$\ddot{u}$neisen
 coefficient (depending on the ratio of the
material thermal expansion coefficient to the specific heat) which is
considered constant with temperature. 

The resonant-mass GW detector NAUTILUS \cite{nau},
 operating at the INFN Frascati
Laboratory, consists of an aluminum alloy 2300-kg bar which can be
cooled to very low temperatures, of the order of 0.1 K, below the
superconducting transition temperature of this alloy, $T_{C}=0.92~$ K
\cite{coccia}.
 The bar is equipped with a capacitive resonant transducer, providing 
the read-out.
Bar and transducer form a coupled
oscillator system with two resonant modes, whose frequencies are
 $906.4$ Hz and $922.0$ Hz.
 The transducer converts the mechanical vibrations into an electrical
signal and is followed by a dcSQUID electronic amplifier. The NAUTILUS data,
recorded with a sampling time of $4.54$ ms, are processed with a
filter \cite{veloce} optimized to detect pulse 
signals applied to the bar, such as those due to a short burst
of GW.

NAUTILUS is equipped with a CR detector system consisting of seven layers of
streamer tubes for a total of 116 counters \cite{rcd}.
 Three superimposed layers, each with an area of $36\rm{m^2}$,
are located over the cryostat (top detector). Four superimposed layers
are set under the cryostat, each with an area of $16.5\rm{m^2}$ 
(bottom detector).
 Each counter measures the
charge, which is proportional to the number of particles. The CR detector 
is able to measure particle density up to $5000$ par./$\rm{m^2}$
 without large saturation
effects and gives a rate of showers in good agreement with the expected
number \cite{rcd,cocco},
 as verified by measuring the particle density in the top detector, 
which is not affected by the interaction in the NAUTILUS bar.\\
 In a previous paper we reported the results of a search for correlation 
between the NAUTILUS data and the data of the CR detector, when for the first 
time acoustic signals generated by CR showers were measured \cite{prl}.
In a further investigation, we found very large NAUTILUS signals 
at a rate much greater than expected \cite{physlet, amaldico}.
(We notice that a GW bar detector, used as particle detector, has
characteristics very different from the usual particle detectors which 
are sensitive only to ionization losses.)
Since the bar temperature
was about 0.14 K, i.e. the aluminum alloy was superconductor,
one could consider some unexpected behaviour due to the transition
 to the normal state along the particle
trajectories. These effects were estimated
\cite{allega,bernard} for type I superconductor (as aluminum).
They are very small and cannot account for our observations, if the
showers include only electromagnetic and hadronic particles.\\

In the present paper, the results of the effect of the CR passage 
on NAUTILUS during the years 2000 and 2001 are reported together with 
comparison with the previous observations. 
During this period NAUTILUS operated at different thermodynamic temperatures.
In 2000, until July, the NAUTILUS bar cryogenic temperature was 0.14 K;
then, between August and December, it was brought at 1.1 K.
In the period 1 March 2001 through 30 September 2001 NAUTILUS operated 
at a temperature of 1.5 K. We proceeded to apply to these
data the same data analysis algorithms used for the previous runs:
coincidence search \cite{physlet, amaldico} and zero threshold search 
\cite{prl}, latter being more efficient for detecting small amplitude signals.\\

\begin{figure}
 \vspace{9.0cm}
\includegraphics{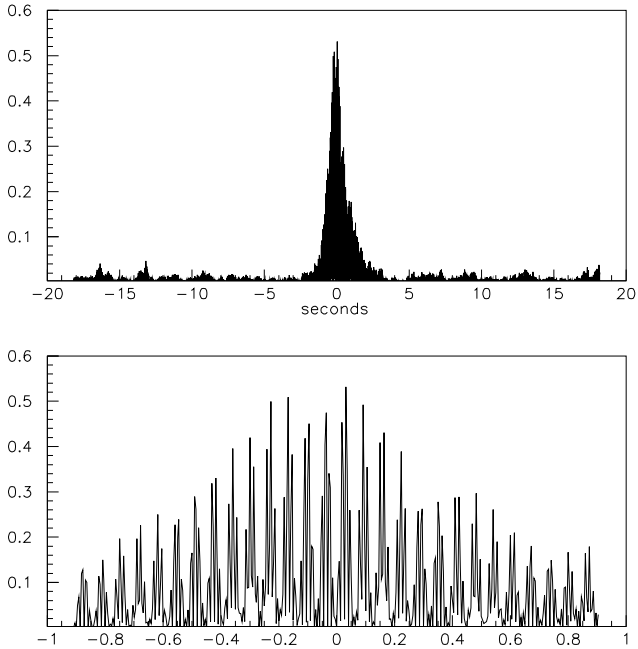}
 \caption{
The NAUTILUS response to the CR shower with particle density
2812 par./$\rm{m^2}$, filtered energy (K) versus time (s),
centered at the CR shower
 arrival time. The lower figure is a zoom of the upper one.
We note the oscillation related to the beating of the two
resonant modes and the decay due to the detector bandwidth,
 $\delta f\sim 0.4$ Hz. 
        \label{big} }
\end{figure}

$Coincidence$ $search.$- The event list employed in the analysis was 
generated by considering only the time periods with noise temperature 
(expressing the minimum detectable innovation)
less than 5 mK, and imposing the amplitude threshold at $\rm{SNR} = 4.4$ on 
the data filtered with an algorithm matched to detect short bursts. 
The threshold value was 
established trough IGEC Collaboration \cite{prodi} for data exchange
among the GW groups to search for coincident events. For each threshold 
crossing we take the maximum
value and its time of occurrence. These two quantities define
the event of the GW detector. The CR shower list was generated by 
considering those events giving a particle 
density $\geq 300$ par./$\rm{m^2}$ in the bottom detector. 
Comparing the two lists, we searched for coincidence 
in a window of $\pm0.1$ s centered at CR arrival time.
The expected number of accidental coincidences 
was experimentally estimated, by means of the 
time shifting algorithm \cite{coin,allegro}. By shifting the events time 
in one of the two data sets by an amount $\delta t$ the number of coincidence
$n(\delta t)$ is determinated. Repeating for N different values of the time 
delay, the expected number of coincidence is 
$\bar{n}=\frac{1}{N-1}\sum{n(\delta t)}$. 
With these criteria, i.e. the temperature noise less than 5 mK, the 
coincidence time window of $\pm{0.1}$ s, and the 
particle density showers larger than 300 par./$\rm{m^2}$,
we found, with  NAUTILUS temperature at 0.14 K, 
12 coincident events on 1998 and 9 coincident events during Feb-Jul 2000.
For both periods, the energy values of the events are concentrated in the 
0.1 K range.
For the remaining part of 2000, with NAUTILUS temperature at 1.1 K,
we found no coincident event. In 2001, with NAUTILUS bar temperature 
at 1.5 K, we found just one coincidence.
We report the result of the analysis and comparison in Table 
\ref{a2000}.
This table affords evidence at about $4~\sigma$ level that the 
observed coincidence rate is related to the bar temperature.
In 2001, the single coincidence event had high NAUTILUS energy, $E\sim~0.5$ K, 
and very large
particle density $M=2812$ par./$\rm{m^2}$ in the bottom detector. 
The response to this CR shower is shown in Fig.\ref{big},
filtered energy versus time centered at the CR shower arrival time.\\
This is the typical response expected for a delta-like excitation acting
on the bar.
To estimate the energy absorbed by the incoming CR shower, we
apply eq.\ref{desudx} to the case of NAUTILUS:
\be
E=7.64~10^{-9}~W^2~f
\label{evw}
\ee
where $E$ is expressed in kelvin units, $W$ in GeV units is
 the energy delivered by
the particle to the bar and $f$  is a geometrical factor of the order of unity.

We get for this event $W\sim 8$TeV. From the data
shown in reference \cite{physlet} (our calculations and experimental data
from the CASCADE collaboration) we expect in 83.5 days of NAUTILUS data
taking about one event with energy greater than 0.1 K
 due to the hadrons, able to deliver to the GW detector an energy of a few TeV.

\begin{table}
\centering
\caption{
Coincidences during the years 1998, 2000 and 2001, using a coincidence
window of $\pm 0.1$ s.
NAUTILUS temperature, duration of the analysis period,
expected number of accidental coincidences $\bar{n}$ and
number of coincidences $n_c$. 
\label{a2000}}
\begin{tabular}{|c|c|c|c|c|c|}
\hline
time period&NAUTILUS&duration&$n_c$&$\bar{n}$&rate\\
&temperature (K)&hours&&&(ev/day)\\
\hline
Sep-Dec 1998&0.14&2002&12&0.47\\
Feb-Jul 2000&0.14& 707& 9&0.42\\
\hline
Total & &2709&21&0.89&$0.178\pm{0.041}$\\
\hline
Aug-Dec 2000&1.1& 118 &0&0.03&\\
Mar-Sep 2001&1.5&2003&1&0.42\\
\hline
Total & &2121&1&0.45&$0.006\pm{0.011}$\\
\hline
\end{tabular}
\end{table}

\begin{figure}
 \vspace{9.0cm}
\includegraphics{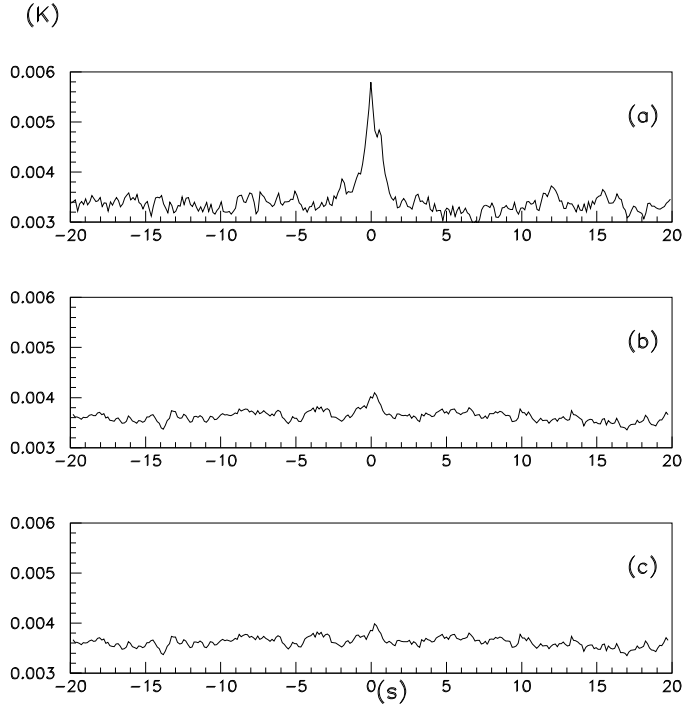}
 \caption{The energy response of NAUTILUS to the CRs passage at zero time.
In fig.(a), we show the average energy (K) vs time for 308 data stretches
detected during 2000, with NAUTILUS bar temperature at 0.14 K.
In fig.(b), the result of the same analysis is shown for 968 data stretches
detected during 2001, with bar temperature at 1.5 K.
The CR showers particle density is larger than $300 \frac{particles}{m^2}$
for the both periods. Excluding from the last data set the event of 
Fig.\ref{big}, the average energy for 967 data stretches 
is shown in the fig.(c).
\label{pina} }
\end{figure}

$Zero$ $threshold$ $search.$- Again we used these data when NAUTILUS noise 
temperature was less than 5 mK and the shower multiplicity was
larger than 300 par./$\rm{m^2}$ in the bottom CR detector.
In correspondence with each CR shower we considered the NAUTILUS filtered data
in a time period of $\pm19$ s centered at the CR arrival time.
With this selection, in 2000, there were 308  data stretches 
corresponding to as many CR showers during a total period of observation 
of 707 non continuous hours.
The selected stretches were superimposed and averaged at the same 
relative time with respect to the arrival time of the CR showers.
The result of this procedure is shown in Fig.\ref{pina}, where we plot the 
averages for each data sample (136.3 ms) versus time, for the 308
CR events with particle density greater than 300 par./$\rm{m^2}$,
graph(a). Several events 
with energy of the order of 0.1 K contribute to the large response at zero 
time, which confirms the results obtained by the data recorded during 1998 
\cite{prl}.
On the same figure, on graph (b), we report the result of the same analysis 
applied to the 968 stretches data of the year 2001, with the  aluminum
 bar cooled to 1.5 K. The major contribution to the signal at zero 
time is due to the single event of Fig.\ref{big}. 
Removing from the data set that event, we obtain graph (c). 
 Comparison shows clearly 
the different response of NAUTILUS in the two time periods.\\
The question arises whether NAUTILUS, operating at temperature T=1.5 K
(in a normal non superconductive status) is sensitive to the CR
 showers as predicted by the thermo-acoustic models.
For a quantitative estimation of a possible effect due to CR we
proceeded as follows. We consider NAUTILUS stretches for the year 2001 
corresponding to CR in various contiguous multiplicity intervals. 
For the stretches of each of the selected multiplicity intervals
we calculate the energy averages over thirty contiguous sampling times
corresponding to 136.3 ms.
At zero delay we take the average at time $0\pm68.2$ ms. We recall
that the beat period in the filtered signal, due to the two resonance modes,
is 64 ms, as we can see from Fig.\ref{big}.  With this averaging procedure 
we avoid the problem of taking either a maximum value or a minimum value,
which are not exactly in phase among the various stretches. By doing so
we get an average value smaller than the maximum by a factor $3.6$,
as we find by numerically averaging the data of Fig.\ref{big}.
For each multiplicity range,
the measured signal (average at time $0\pm68.2$ ms) is compared
with the signal we expect due to the electromagnetic component of the
shower. The theoretical value is given by \cite{physlet}
\be
E_{th}=\Lambda^2\cdot4.7~10^{-10}~K
\label{lambda}
\ee
where $\Lambda$ is the number of secondaries through the bar.
The measured multiplicity
might be affected by a systematic error of the order of $\pm25\%$ 
\cite{physlet}.
As an estimate of the background we take the average energy during the
periods from - 4000 to - 3000 sampling times (from - 18.18 s to - 13.63 s)
and from 3000 to 4000 samplings times (from 13.63 s to 18.18 s), for
a total time period of 2000 sampling times, 9.088 seconds.

\begin{figure}
 \vspace{9.0cm}
\includegraphics{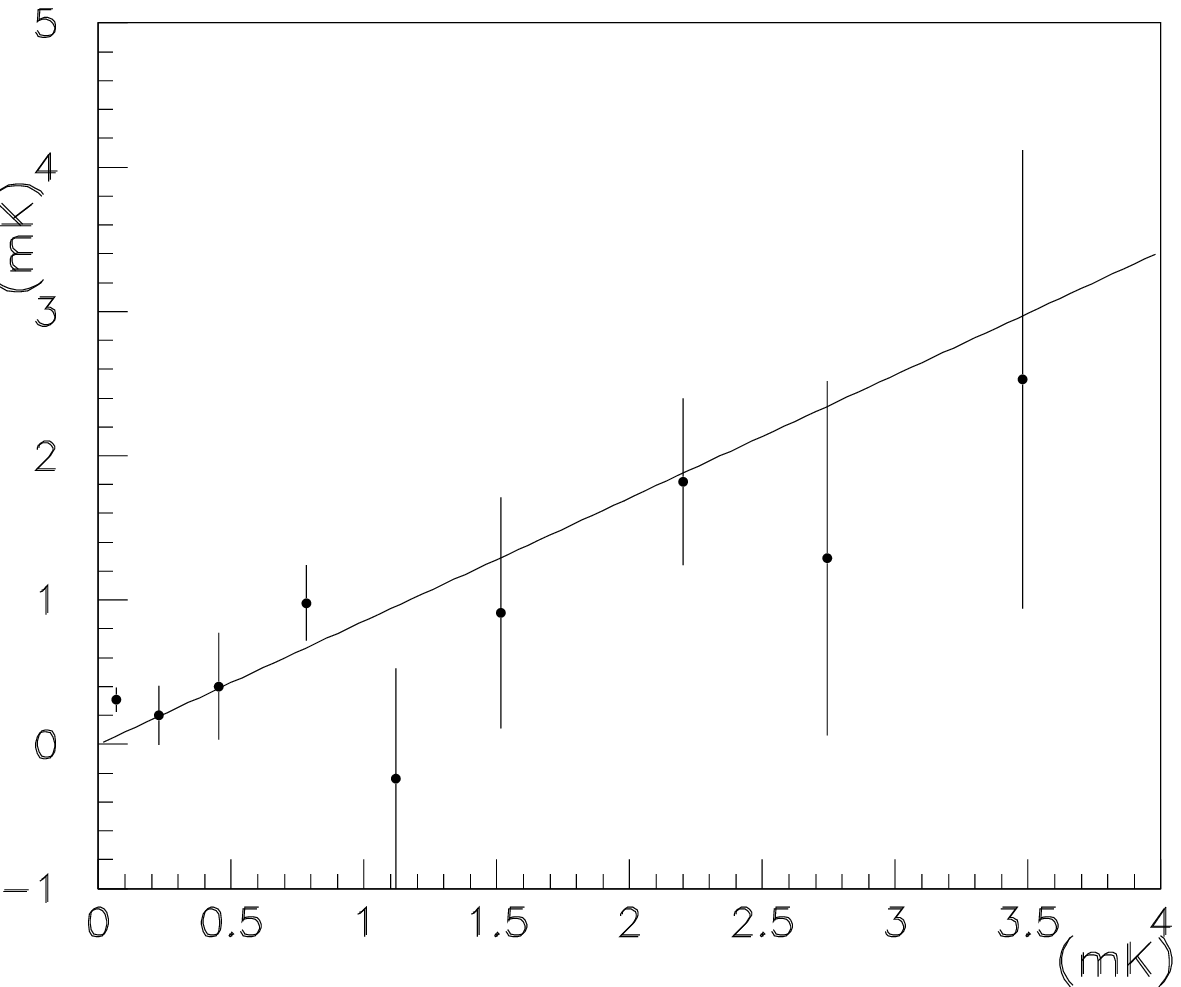}
 \caption{Experimental signal versus the expected signal due to the 
electromagnatic component of the CR shower (see text and 
Table \ref{diffetavola}). The straight line is at least square fit and 
the vertical bars indicate statistical errors ($\pm$ one standard deviation).
\label{diffe}}
\end{figure}

In Fig. \ref{diffe} we show the difference in mK units between the
average energy at zero time delay and the background 
versus the expected signal due to the electromagnetic component
of the CR showers.
The straight line is a least square fit through the origin 
and the vertical bars 
indicate statistical errors ($\pm$ one standard deviation).
The $\chi^2$ calculated for a null hypothesis (signal=background) gives
$\chi^2=42.4$ with 9 degrees of freedom for a probability of $2.8~10^{-6}$.
The slope of the straight line has value $0.85\pm0.13$. If we take into account
the systematic error on the experimental value of $\Lambda$ 
($\sim \pm~25\%$) and the error on the calibration of the NAUTILUS event
energy, of the order of $10\%$, we get for the slope 
$0.85\pm0.16\pm0.42$, showing a good agreement with the 
thermo-acoustic model. The $\chi^2$ calculated for
the hypothesis that the individual data be along the straight line
is $\chi^2=13.3$ for a probability of $0.10$.
\begin{table}
\centering
\caption{
The average NAUTILUS signal $\rm{E_{obs}}$, and its standard deviation,
vs the multiplicity of CR events.
multiplicity selections. Also indicated are the number of stretches for each
selection and the difference between the
signal at zero delay and the background, with its standard deviation.
The theoretical values are calculated with eq.\ref{lambda}, (valid for the 
electromagnetic component of the shower)
divided by 3.6, using the measured particle density in the lower part 
of CR detector and taking the average. 
The big event of Fig.\ref{big} has been excluded from the last row.}
\vskip 0.1 in
\begin{tabular}{|c|c|c|c|c|c|c|}
\hline
$\frac{\rm_{particles}}{\rm{m^2}}$&number of& $E_{obs}$&$\sigma_{E_{obs}}$&
$E_{obs}-bkg$&$\sigma_{E_{obs}-bkg}$&$E_{th}$\\
&stretches&mK&mK&mK&mK&mK\\
\hline
 300 -  600&  688& 3.690& 0.085& 0.310& 0.085& 0.069\\
 600 -  900&  138& 3.67& 0.21& 0.20& 0.21& 0.228\\
 900 - 1200&   63& 3.91& 0.37& 0.40& 0.37& 0.453\\
1200 - 1500&   34& 4.10& 0.26& 0.98& 0.26& 0.783\\
1500 - 1800&   16& 3.35& 0.77& -0.24& 0.79& 1.119\\
1800 - 2100&    9& 4.76& 0.80& 0.91& 0.84& 1.517\\
2100 - 2400&   11& 4.75& 0.58& 1.82& 0.60& 2.200\\
2400 - 2700&    3& 4.4& 1.2& 1.3& 1.4& 2.744\\
2700 - 3000&    5& 6.2& 1.6& 2.5& 1.7& 3.478\\                   

\hline
\end{tabular}
\label{diffetavola}
\end{table} 

$Conclusions.$ Comparing the previous and the present 
measurements, two different 
behaviours of the  aluminum bar detector are noticed, with evidence at
4 $\sigma$ level.\\
In the run with the bar temperature above the superconductive transition 
we find a 
result in good agreement with the theoretical predictions of the 
thermo-acoustic model.
These measurements are a record for the GW detectors, as signals of
 the order of $10^{-4}$ K, corresponding to $10^{-8}$ eV, were 
extracted from noise.
The unexpected behaviour of NAUTILUS noticed in the previous runs
\cite{physlet,amaldico}, i.e.
very large signals at a rate higher than expected, occurs only 
at ultracryogenic temperatures. The observed phenomenology suggests a larger 
efficiency in the mechanism of conversion of the particles energy into
the vibrational mode energy, at least for some type of particles,
when the  aluminum bar is in the superconductive status.

$Acknowledgements.$
We thank F. Campolungo, G. Federici, R. Lenci, G. Martinelli, E. Serrani,
 R. Simonetti and F. Tabacchioni for precious technical assistance.
We also thank dr R. Elia for her contribution.

%
%

%

\begin{thebibliography}{99}

\bibitem{prl}P. Astone et al. (ROG Coll.), Phys. Rev. Lett. {\bf84}, 14 (2000).
\bibitem{physlet}P. Astone et al. (ROG Coll.), Phys. Lett. B {\bf499}, 16 (2001)
\bibitem{amaldico}P. Astone et al. (ROG Coll.),
Proc. of "4th E. Amaldi Conference", Perth, 2001.
\bibitem{beron}B.L. Beron and R. Hofstander, Phys. Rev. Lett. {\bf23}, 184 (1969).
\bibitem{beron1}
B.L. Beron, S.P. Boughn, W.O. Hamilton, R. Hofstander, T.W. Tartin, IEEE
Trans. Nucl. Sci. {\bf17}, 65 (1970).
\bibitem{strini}
A.M. Grassi Strini, G. Strini and G.Tagliaferri, J. Appl. Phys. {\bf51}, 849 (1980).
\bibitem{allega}A.M. Allega and N. Cabibbo, Lett. Nuovo Cimento {\bf83}, 263 (1983).
\bibitem{bernard}
C. Bernard, A. De Rujula and B. Lautrup, Nucl. Phys. B {\bf242}, 93 (1984).
\bibitem{deru}A. De Rujula and S.L. Glashow, Nature 312, {\bf734} (1984).
\bibitem{amaldi}E. Amaldi and G. Pizzella, Il Nuovo Cimento {\bf9}, 612 (1986).
\bibitem{bari}G. Liu and B. Barish, Phys. Rev. Lett. {\bf61}, 271 (1988).
\bibitem{nau}
P. Astone P. et al. (ROG Coll.), Astropar. Phys. {\bf7}, 231  (1997).
\bibitem{coccia}E. Coccia and T. Niinikoski, J. of Physics E {\bf16}, 695  (1983).
\bibitem{veloce}P. Astone et al. Il Nuovo Cimento {\bf20}, 9 (1997).
\bibitem{prodi}G. Prodi et al. (IGEC Coll.),
Proc. of 4th Gravitational Wave Data Analysis Workshop
(GWDAW 99), Rome, Italy, 2-4 Dec 1999.
\bibitem{rcd}E. Coccia et al. Nucl. Instr. and Methods  A {\bf335}, 624 (1995).
\bibitem{cocco}
G. Cocconi, Encyclopedia of Physics, ed. by S. $\rm{Fl\ddot{u}gge}$,
Vol. 46, p. 228 (1961).
\bibitem{coin}P. Astone and G. Pizzella,
Int.Rep. INFN - LNF, LNF-95-{\bf003(P)}, (1995).
\bibitem{allegro}P. Astone et al. (ROG Coll.), Phys. Rev. D, 
{\bf59}, 122001 (1999).
\bibitem{sih}F. Sihoan et al. J. Phys. G{\bf3}, 8 (1977).
\bibitem{peter}
J. Chiang, P. Michelson and J. Price,  Nucl. Instr. and Meth. A {\bf311}, 363
(1992).
\bibitem{heck}
D. Heck et al. Report FZKA 6019, Forschungszentrum Karlsruhe (1998).
\bibitem{ambrosio}M. Ambrosio  et al. (MACRO Coll.),
 Phys. Rev. D, {\bf56}, 1418 (1997).
\bibitem{hora}
J.R. Horandel et al. ICRC Cosmic Ray Conference, Salt Lake City, 
{\bf1}, 337 (1999).
\bibitem{desalvo}R. Desalvo, Proc. Amaldi Conference on Gravitational
Waves, Ed. E. Coccia, G. Pizzella and G. Veneziano, CERN July 1997,
World Scientific
\bibitem{fitz} E.R. Fitzgerald, Nature, {\bf252}, 638 (1974).
\bibitem{witten}E. Witten, Phys. Rev. D {\bf30}, 272 (1984).
\bibitem{nucleariti}P. Astone et al. (ROG Coll.), Phys. Rev. D {\bf47}, 10, 4770 (1993).

\end{thebibliography}
\end{document}